%% file: main.tex
\documentclass[submission,copyright]{eptcs}
\providecommand{\event}{FTSCS 2012} 
\usepackage{breakurl}             
\usepackage{graphicx}
\usepackage{amsmath}
\usepackage{amssymb}
\usepackage{subfigure}
\usepackage{url}
\usepackage{wrapfig}

\input{macro.tex}

\title
{
    {\modediagram}: A Mode Diagram Modeling Framework for \\ Periodic Control Systems
}

\author{
Zheng Wang, Geguang Pu\thanks{Corresponding Author},  Jianwen Li, Jifeng He
\institute{Shanghai Key Laboratory of Trustworthy Computing}
\institute{East China Normal University}
\email{\{wangzheng,ggpu,jifeng\}@sei.ecnu.edu.cn}
\and
Shenchao Qin
\institute{University of Teesside}
\email{s.qin@tees.ac.uk}
\and
Kim G. Larsen
\institute{Aalborg University of Denmark}
\email{kgl@cs.aau.dk}
\and
Jan Madsen
\institute{Technical University of Denmark}
\email{jan@imm.dtu.dk}
\and
Bin Gu
\institute{Beijing Institute of Control Engineering}
\email{gubin88@yahoo.com.cn}
}

\def\titlerunning{{\modediagram}: A Mode Diagram Modeling Framework}
\def\authorrunning{Zheng WANG et al. }

\begin{document}
\maketitle

\begin{abstract}
Periodic control systems  used in spacecrafts and automotives are usually period-driven and can be decomposed into different modes with each mode representing a system state observed from outside. Such  systems may also involve  intensive computing in their modes.
Despite the fact that such control systems are widely used in the above-mentioned safety-critical embedded  domains, there is lack of domain-specific formal modeling languages for such systems in the relevant industry.
To address this problem, we propose a formal visual modeling framework called  {\modediagram} as a concise and precise way to specify and analyze such systems. To capture the temporal properties of  periodic control systems, we provide, along with {\modediagram},  a property specification language based on interval logic for the description of concrete temporal requirements the engineers are concerned with. The statistical model checking technique can then be used to verify the {\modediagram} models against the desired properties. To demonstrate the viability of our approach, we have applied our modeling framework to some real-life case studies from industry and helped detect two design defects for some spacecraft control system.
\end{abstract}


\input intro.tex

\input lang.tex

\input property.tex

\input verification.tex

\input related.tex

\section*{Acknowledgement}

WANG Zheng and GU Bin are partially supported by Projects NSFC
No.\ 90818024 and NSFC No.\ 91118007. PU Geguang is partially supported
by NSFC Projects No.\ 61061130541 and No.\ 61021004. Jianwen Li is partially supported by Shanghai Knowledge Service Platform for Trustworthy Internet of Things (No. ZF1213). Jifeng He is partially supported by 973 Project No. 2011CB302904. Shengchao Qin was supported in part by EPSRC project EP/G042322.

\bibliographystyle{eptcs}
\bibliography{modechart}
\end{document}

%% file: macro.tex
\newcommand{\hide}[1]{{}}
\def \modediagram {{\textsf{MDM}}}
\def \State{\textit{State}}
\def \MD {\textit{md}}
\def \mlist{\textit{mlist}}
\def \TopModes {\emph{TopModes}}
\def \Contains {\textit{Contains}}
\def \Trans {\textit{Trans}}

\def \PBegin   {{\textit{Begin}}}
\def \PExecute {{\textit{Execute}}}
\def \PEnd     {{\textit{End}}}
\def \CFGStart {{\textit{Start}}}
\def \CFGExit  {{\textit{Exit}}}
\def \sampling {{\mathsf{sampling}}}
\def \execute  {{\textit{execute}}}
\def \failure  {{\textit{failure}}}

\newcommand{\submode}[2]{\mathsf{sub\_mode}(#1, #2)}
\newcommand{\supermodes}[2]{\mathsf{super\_modes}(#1, #2)}
\newcommand{\upmodes}[3]{\mathsf{up\_modes}(#1, #2, #3)}
\newcommand{\outs}[1]{\mathsf{outs}(#1)}
\newcommand{\CHOP} {{^{\frown}}}

\def \kwVar {\textit{Var}}
\def \kwMode {\textit{Mode}}
\def \kwModule {\textit{Module}}
\def \kwname {\textit{name}}
\def \kwperiod {\textit{period}}
\def \kwinitial {\textit{initial}}
\def \kwBody {\textit{Body}}
\def \kwTransition {\textit{Transition}}
\def \kwguard {\textit{guard}}
\def \kwpriority {\textit{priority}}
\def \kwtarget {\textit{target}}
\def \kwsource {\textit{source}}
\def \kwCFG {\textit{CFG}}

\def \kwSystem {\textit{md}}
\def \kwConst {\textit{Const}}
\def \kwSExpr {\textit{SExpr}}
\def \kwBTerm {\textit{BTerm}}
\def \kwBExpr {\textit{BExpr}}
\def \kwIExpr {\textit{IExpr}}
\def \kwGTerm {\textit{GTerm}}
\def \kwGuard {\textit{guard}}
\def\sigmasq{\ensuremath{\sigma_1{\cdot}...{\cdot}\sigma_n}}

\def \SA {\emph{A}}

\newcommand{\period}[1]{\mathsf{period(#1)}}
\newcommand{\isInitial}[1]{\mathsf{is\_initial(#1)}}
\newcommand{\CFG}[1]{\mathsf{CFG(#1)}}

\newcommand{\priority}[1]{\mathsf{prio(#1)}}
\newcommand{\guard}[1]{\mathsf{guard(#1)}}
\newcommand{\target}[1]{\mathsf{target(#1)}}
\newcommand{\source}[1]{\mathsf{source(#1)}}

\newcommand{\True}{\mathsf{\mathsf{true}}}
\newcommand{\False}{\mathsf{\mathsf{false}}}
\newcommand{\tvalue}{\textit{tt}}
\newcommand{\fvalue}{\textit{ff}}
\newcommand{\timestamp}{\textit{timestamp}}

\newcommand{\Int}{\textit{I}nt}

\newcommand{\Terms}{\textit{Terms}}
\newcommand{\Formulas}{\textit{Formulas}}
\newcommand{\Intv}{\textit{Intv}}

\newcommand{\after}{\mathsf{after}}
\newcommand{\duration}{\mathsf{duration}}

\newcommand{\runa}[1]{\mbox{\textsc{\protect{(#1})}}}

\newcommand{\infrulesL}[3]
           {\parbox{#1mm}{$$ \frac{#2}{#3}\hspace{.5cm}$$}}

\def \stmts    {\textit{stmts}}
\def \aStmt    {\textit{aStmt}}
\def \pStmt    {\textit{pStmt}}
\def \cStmt    {\textit{cStmt}}
\newcommand{\CALL}{\mathsf{call}}
\newcommand{\WHILE}{\mathsf{while}}
\newcommand{\DO}{\mathsf{do}}
\newcommand{\IF}{\mathsf{if}}
\newcommand{\THEN}{\mathsf{then}}
\newcommand{\ELSE}{\mathsf{else}}

\newcommand{\SKIP}{\mathsf{skip}}

%% file: intro.tex
\section{Introduction}

\hide{Periodic control systems are widely used in embedded computing area, for instance, spacecraft control system and automotive control system etc. Such systems are usually driven by a global or local clock, which triggers the behavior of the system in periods. Another feature of these systems is that they can be decomposed into several different modes from the system view.  Each mode actually represents an important state of the system. Further more, a mode can also be divided into several sub-modes. When a period ends, the system may switch from the current mode to another mode if the transformation condition holds. For instance, China Academy of Space Technology (CAST) spends great efforts in designing and developing embedded software for spacecrafts. The periodic control system is a key for the spacecrafts, which has the following features:}%

The control systems that are widely used in safety-critical embedded domains, such as spacecraft control and automotive control, usually reveal periodic behaviors. Such {\em periodic} control systems share some interesting features and characteristics:
\begin{figure*}[t]
    \centering
    \includegraphics[width=0.98\textwidth]{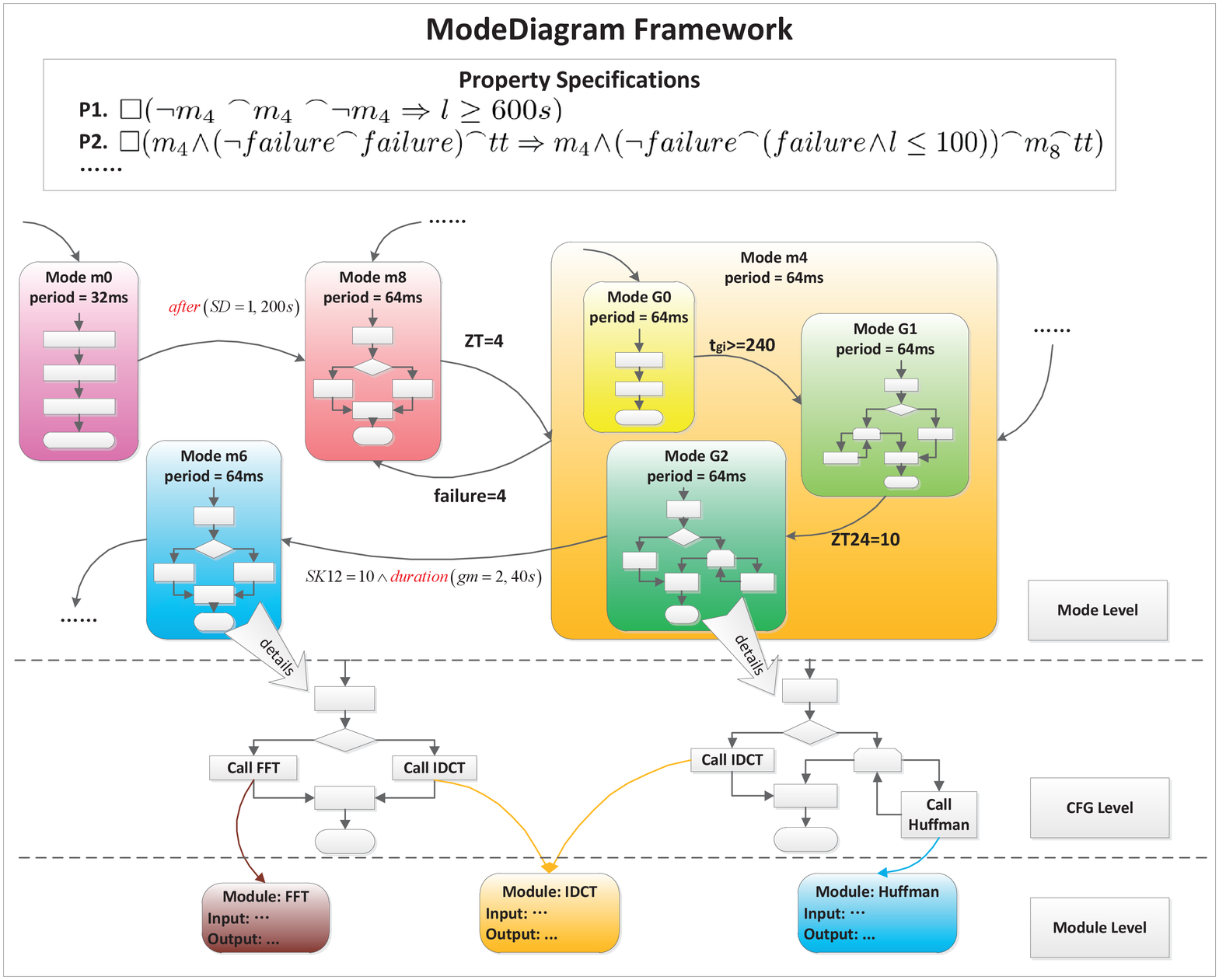}\\
    \vspace*{-3mm}
    \caption{ {\modediagram}: An (Incomplete) Example }\label{fig:mcoverall}\vspace*{-3mm}
\end{figure*}

\begin{itemize}
    \item They are {\em mode-based}.  A periodic control system is usually composed of a set of modes, with each mode representing an important state of the system. Each mode either contains a set of sub-modes or performs controlled computation periodically.
    \item They are {\em computation-oriented}. In each mode,  a periodic control system may perform control algorithms involving complex computations. For instance, in certain mode, a spacecraft control system may need to process intensive data in order to determine its space location.

    \item They behave {\em periodically}.  A periodic control system is reactive and may run for a long time. The behavior of each mode is regulated by its own period. That is, most computations are performed within a  period and may be repeated in the next period if mode switch does not take place. A mode switch may only take place at the end of a period under certain conditions.

\end{itemize}

Despite the fact that periodic control systems have been widely used in areas such as spacecraft control, there is a lack of a concise and precise domain specific formal modeling language for such systems. In our joint project with China Academy of Space Technology (CAST), we have started with several existing modeling languages but they are either too complicated therefore require too big a learning curve for domain engineers, or  are too specific/general, therefore require non-trivial restrictions or extensions.  This motivates us to propose  a new formal but lightweight modeling language that matches exactly the need of the domain engineers,  the so-called  \underline{M}ode \underline{D}iagram \underline{M}odeling framework ({\modediagram}).

Although the proposed modeling notation {\modediagram} can be regarded as a variant of Statecharts \cite{statechart}, it has been specifically designed to cater for the domain-specific need in modeling periodic control systems. We shall now use an example  to illustrate informally the {\modediagram} framework, and leave the formal syntax and semantics to the next section. As shown in Fig~\ref{fig:mcoverall}, the key part of an {\modediagram} model is the collection of modes given in the mode level.  Each mode has a period, and the periods for different modes can be different. A mode can be nested and the transitions between modes or sub-modes may take place. A transition is enabled  if the associated guard is satisfied. In {\modediagram}, the transition guards may involve complex temporal expressions. For example, in the transition from mode {\sf G2} to mode {\sf m6}, in addition to the condition $\textsf{SK12=10}$, it also requires that the condition $\textsf{gm=2}$ has held for 40s, as captured by the $\mathsf{duration}$ predicate.

An {\modediagram} model is presented hierarchically. A mode that does not contain any sub-mode (termed a {\em leaf} mode) contains a control flow graph (CFG) encapsulating specific control algorithms or computation tasks. The details of CFGs are given in the CFG level. The CFGs may refer to modules (similar to procedures in conventional languages) details of which are given in the module level.

To support formal reasoning about {\modediagram} models, we also provide a property specification language inspired by an interval-like calculus~\cite{IntervalCalculus}, which facilitates the capture of temporal properties  system engineers may be interested in. Two example properties are listed in Fig~\ref{fig:mcoverall}. The property {\sf P1} says that ``whenever the system enters the {\sf m4} mode, it should stay there for at least 600s''. The formal details of the specification language is left to a later section.

To reason about whether an {\modediagram} model satisfies  desired properties specified by system engineers using the property specification language, we employ statistical model checking techniques~\cite{SMCYounes05,SMCYounesS02}. Since {\modediagram} may involve complex non-linear computation in its control flow graph, complete verification is undecidable. Apart from incompleteness, statistical model checking can verify  hybrid systems efficiently \cite{StatHeterogeneous}. Our experimental results on real-life cases have demonstrated that statistical model checking can help uncover  potential defects of {\modediagram} models.

In summary, we have made the following contributions in this paper:

\begin{itemize}\setlength{\itemsep}{0pt}
    \item  We propose a novel visual formal modeling notation {\modediagram}  as a concise yet precise modeling language for  periodic control systems. Such a notation is inspired from the industrial experiments of software engineers.
    \item  We present a formal semantics for {\modediagram} and a property specification language to facilitate the verification process.
    \item We develop a new statistical model checking algorithm   to verify   {\modediagram} models against various temporal properties. Some real-life case studies have been carried out to demonstrate the effectiveness of the proposed framework.  Furthermore,  the design defects of a real spacecraft control system are discovered by our approach.
\end{itemize}

The rest of this paper is organized as follows. Section~\ref{sec:modechart} presents the formal syntax and semantics of {\modediagram}. Section~\ref{sec:specification} introduces our interval-based property specification language and its semantics. The statistical model checking algorithm for {\modediagram} is developed in Section~\ref{sec:statistical}, followed by related work and concluding remarks.

%% file: lang.tex
\section{The {\modediagram} Notations}\label{sec:modechart}

Before developing the formal model of {\modediagram}, we will begin by giving its informal description. An  {\modediagram} model is composed by several modes, variables used in the mode, and mode transitions specifying the mode switch relations. A mode essentially refers to the state of the system which can be observed from outside. The mode body can be either a Control Flow Graph (CFG), which prescribes the computational tasks the system can perform in every period, or several other modes as sub-modes. If the mode has sub-modes, when the system is in this mode, it should be in one of its sub-modes. We say that the mode is a leaf mode if its mode body is a control flow graph. A leaf mode usually encapsulates the control algorithms involving complicated computations. The CFG in a leaf mode follows the standard notation, which contains assignment, conditional and loop. It also supports function units similarly to the ones in programming languages.

\subsection{The Syntax of {\modediagram}}

\begin{figure}[t]
    \centering
    \begin{tabular}{c}
    \begin{tabular}{cc}
    $
    \begin{small}
        \begin{array}{rcl}
          \kwSystem     & ::= & (\kwVar^{+}, \kwMode^{+}, \kwModule^{+})  \\
          \kwMode       & ::= & (\kwname, \kwperiod, \kwinitial, \\
                        &     & \hspace*{2mm} \kwBody, \kwTransition^{+})  \\
          \kwBody       & ::= & \kwMode^{+} \mid \kwCFG \\
          \kwTransition & ::= & (\kwsource,\kwguard, \kwpriority, \kwtarget) \\
          \kwModule     & ::= & (\kwname, V_I, V_O, CFG)
        \end{array}
    \end{small}
    $
    &
    $
    \begin{small}
        \begin{array}{rcl}
            \kwSExpr & ::= & \kwConst \mid \kwVar \mid f^{(n)}(\kwSExpr \ldots) \\
            \kwBTerm & ::= & \True \mid \False \mid p^{(n)}(\kwSExpr \ldots) \\
            \kwIExpr & ::= & (\after \mid \duration) ( \kwBTerm, \kwSExpr ) \\
            \kwGTerm & ::= & \kwIExpr \mid \kwBTerm \\
            \kwBExpr & ::= & \hspace*{1mm} \kwBTerm \mid \neg \kwBExpr \\
                     &     & \mid \kwBExpr \vee \kwBExpr \mid \kwBExpr \wedge \kwBExpr \\
            \kwGuard & ::= & \hspace*{1mm} \kwGTerm \mid \neg \kwGuard \\
                     &     & \mid \kwGuard \vee \kwGuard\mid \kwGuard \wedge \kwGuard \\
        \end{array}
    \end{small}
    $
    \\
    (a){\modediagram} & (b) Expressions and Guards
    \end{tabular} \\
    \\
    \begin{tabular}{c}
    $
            \begin{array}{rcl}
                 CFG & ::= &  \stmts \\
              \stmts & ::= &  \pStmt \mid \cStmt\\
              \pStmt & ::= &  \aStmt \mid \CALL ~ name \mid \SKIP ~ |\\
              \aStmt & ::= &  x := \kwSExpr \\
              \cStmt & ::= &  \stmts ;~\stmts \mid  \WHILE ~ \kwBExpr ~ \DO ~ \stmts \mid  \\
                     &     &  \IF ~ \kwBExpr ~  \THEN ~ \stmts ~ \ELSE ~ \stmts
            \end{array}
        $
    \\
    (c) CFG
    \end{tabular}
    \end{tabular}
    \caption{The Syntax of {\modediagram}}\label{fig:mcsyntax}
\end{figure}

We briefly list its syntactical elements in Fig.~\ref{fig:mcsyntax}(a). An {\modediagram} is composed of a list of modes ($\kwMode^+$) and modules ($\kwModule^{+}$), as well as a list of variables ($\kwVar^+$) used in those modes and modules.

Intuitively, a mode  refers to a certain state of the system which can be observed from outside. A mode has a name, a period, a body and a list of transitions. For simplicity, we assume all mode names are distinct in an MDM model. The mode period (an integer number) is used to trigger the periodic behavior of the  mode. The $\kwinitial$ denotes a mode is an initial mode or not. The mode body can be composed of either a control flow graph (CFG), prescribing the computational tasks the system can perform in the mode in every period, or a list of other modes as the immediate sub-modes of the current mode. If a mode has sub-modes, when the control lies in  this mode, the control should also be in one of the sub-modes. A leaf mode does not have sub-modes, so  its body contains a CFG. A mode is either a leaf mode, or it directly or indirectly has  leaf modes as its sub-modes. A mode is called top mode if it is not a sub-mode of any other mode. The CFG in a leaf mode is the standard control flow graph, which contains nodes and structures like assignment, module call, conditional and loop. It also supports  function units like the ones in conventional programming languages. The syntax of CFG is presented in Figure~\ref{fig:mcsyntax}(c).

A module encapsulates computational tasks as its CFG. $V_I$ specifies the set of variables used in the CFG, while $V_O$ is the set of variables modified in the CFG. A module can be invoked by some modes or other modules. As a specification for embedded systems, recursive module calls are forbidden.

A transition (from $\kwTransition^+$) specifying a mode switch from  one mode to another is represented as a quadruple, where the first element is the name of the source mode, the second specifies the transition condition,  the third  is the priority of the transition and the last element is the name of the target mode. The {\modediagram} supports mode switches at different levels in the mode-hierarchy. The transition condition (i.e. $\kwguard$) is defined in Fig.~\ref{fig:mcsyntax}(b). A state expression can be either a constant, a variable, or a real-value function on state expressions. A boolean term is either a boolean constant, or a predicate on state expressions. There are two kinds of interval expressions, $\mathsf{after}$ and $\mathsf{duration}$. These interval expressions are very convenient to model system behaviors related with past states. A guard term can be either an interval expression, or a boolean term. A guard is the boolean combination of guard terms. To ensure that the mode switches be deterministic, we require that the priority of a transition has to be different from the others in the same mode chain:
\[
    \forall m \in \kwMode \cdot \forall t_1, t_2 \in \outs{\supermodes{\MD}{m}} \cdot
    t_1 \ne t_2 \Rightarrow \priority{t_1} \ne \priority{t_2}
\]
The functions $\supermodes{\MD}{m}$ and $\outs{\mlist}$ will be defined later.

\subsubsection{Auxiliary Definitions}

Given an {\modediagram}
$
\begin{small}
\kwSystem ::= (\kwVar^{+}, \kwMode^{+}, \kwModule^{+})
\end{small}
$,
we introduce two auxiliary relations:\\
$Contains(\kwSystem) \subseteq {\kwMode}s \times {\kwMode}s$ for mode-subsume relation and\\
$Trans(\kwSystem) \subseteq {\kwMode}s \times \Int \times \kwGuard \times {\kwMode}s$ for mode-switch  relation.

Given a mode $m = (n, per, ini, b, tran)$ and a transition $t = (m, g, pri, m')$, we define these operations/predicates:
\[
    \begin{array}{llll}
        \period{m} = per   ~&~ \isInitial{m} = ini ~& ~\CFG{m} = b   ~&~                 \\
        \priority{t} = pri ~&~ \guard{t} = g       ~& ~\source{t}=m  ~&~ \target{t} = m'  \\
    \end{array}
\]
We also define the following auxiliary  functions:
\[
    \begin{array}{l}
        \supermodes{\MD}{m} \triangleq \langle m_1, m_2, \ldots, m_k \rangle, \text{ where }\\
        \hspace*{5mm} m_k = m \wedge m_1 \in {\TopModes}(\MD) \wedge \forall 1 {<} i {\le} k \cdot (m_{i-1}, m_i) \in \Contains(\MD) \\
        \hspace*{5mm} \text{and } m{\in}{\TopModes}(\MD) \triangleq  m {\in} {\kwMode}s(\MD) {\wedge} \neg\exists m'\hide{ {\in} {\kwMode}s}\cdot ( m', m ) { \in} \Contains(\MD) \\
        \\
       \upmodes{\MD}{m}{k} \triangleq \{ m_i \mid m_i \in \supermodes{\MD}{m} \wedge \mod(k, \frac{\period{\mathit{m_i}}}{\period{\mathit{m}}}) = 0 \} \\
       \\
       \submode{\MD}{m} \triangleq m', \text{ where } (m, m') \in \Contains(\MD) \wedge \isInitial{\mathit{m'}} \\
       \\
       \outs{\MD,\mlist} \triangleq \bigcup_{m \in \mlist}\{t \mid t \in \Trans(\MD) \wedge \source{\mathit{t}}=m\}
    \end{array}
\]
The function $\supermodes{\MD}{m}$  retrieves a sequence of modes from a top mode to $m$ using the $\Contains$ relation. The set ${\TopModes}(\MD)$ consists all the modes which are not sub-modes of any other mode. The function $\upmodes{\MD}{m}{k}$ returns those modes in $\supermodes{\MD}{m}$ whose periods are consistent with the period count $k$. An {\modediagram} requires that the period of a mode should be equal to or multiple to the period of its sub-modes. The function $\submode{\MD}{m}$ returns the initial sub-mode for a non-leaf node $m$, and the predicate $\isInitial{\mathit{m'}}$ means that the sub-mode $m'$ is the initial sub-mode in its hierarchy. The function $\outs{\mlist}$ returns all outgoing transitions from modes in $\mlist$.

\subsection{The Semantics}\label{subsec:semantics}

In order to precisely analyze the behaviors of {\modediagram}, for instance, model checking of {\modediagram} , we need its formal semantics. In this section, we present the operational semantics for {\modediagram}.

\begin{table}[t]
    \[
    \begin{small}
    \begin{array}{lcl}
      \sigmasq \models b                          & \Leftrightarrow & \sigma_n \models b \\
      \sigmasq \models \neg g                     & \Leftrightarrow & \neg(\sigmasq \models g)  \\
      \sigmasq \models g_1 \vee g_2               & \Leftrightarrow & \sigmasq \models g_1 \text{ or } \sigmasq \models g_2 \\
      \sigmasq \models g_1 \wedge g_2             & \Leftrightarrow & \sigmasq \models g_1 \text{ and } \sigmasq \models g_2 \\
      \sigmasq \models \mathsf{duration}(b, l)    & \Leftrightarrow & \sigma_n(l) = \nu \wedge  \exists i {<}n \cdot ( \sigma_i(ts) {+} \nu\leq \sigma_n(ts) \wedge \\
                                                                  &                   &
       \sigma_{i{+}1}(ts) {+} \nu \geq \sigma_n(ts) \wedge       \forall i {\leq} j {\leq} n \cdot \sigma_j(b) = \True)  \\
      \sigmasq \models \mathsf{after}(b, l)       & \Leftrightarrow & \sigma_n(l) = \nu \wedge       \exists i {<} n \cdot (   \sigma_i(ts) {+}\nu \leq \sigma_n(ts) \wedge\\
                                                                  &                   &
 \sigma_{i{+}1}(ts) {+} \nu \geq \sigma_n(ts)) \wedge      \sigma_i(b) = \True) 
    \end{array}
    \end{small}
    \]
    \caption{The Interpretation of Guards}\vspace*{-6mm}
    \label{tbl:evalie}
\end{table}

\subsubsection{Configuration}

The configuration in our operational semantics is represented as $( \MD, m, l, pc, k, \Sigma )$, where
\begin{itemize}
    \item $\MD$ is the \modediagram, and $m$ is the  mode  the system control currently lies in.
    \item $l \in \{\PBegin, \PExecute, \PEnd\}$ specifies the system is in the beginning, middle, or end of a period.
    \item $pc \in \mathcal{L}$, where $\mathcal{L} = \mathcal{N} \cup \{\CFGStart, \CFGExit, \bot\}$ is the program counter to execute the control flow graph. $\mathcal{N}$ is used to represent the nodes in control flow graphs and $\CFGStart$, $\CFGExit$ denote the start and exit locations of a control flow graph respectively. If the current mode is not equipped with any flow graph, we use the symbol $\bot$ as a placeholder.
    \item The fourth component $k$ records the count of periods for the current mode. If the system switches to another mode, it will be reset to $1$. The period count is used to distinguish whether a super-mode of the current mode is allowed to check its mode switch guard.
    \item $\Sigma$ is a list of states of the form $\Sigma' \cdot \sigma$, where $\sigma$ denotes the current state ($\sigma\in \State \triangleq \textit{Vars}{\rightarrow}\mathbb{R}$) and $\Sigma'$ represents a history of states.

\end{itemize}

\noindent{\bf Guards\quad}The evaluation of a transition guard may depend on the current state as well as some historical states. Table~\ref{tbl:evalie} shows how to interpret a guard in a given sequence of states. The symbol $ts$ is the abbreviation of the variable $\timestamp$. The guard $\mathsf{duration}(b,l)$ evaluates to $\True$ if the boolean expression $b$ has been  $\True$  during the time interval $l$ up to the current moment. The guard  $\mathsf{after}(b,l)$ evaluates to $\True$ if  the boolean expression $b$ was $\True$  the time interval $l$ ago. In this table, $b$ is a pure boolean expression
without interval expressions and $l$ is a state expression.

\subsubsection{Operational Rules}
\label{subsec:inferencerules}

\begin{table}[t]
    \centering
    \[
    \begin{small}
    \begin{array}{cc}
        \runa{enter} & \infrulesL{120}
                        {
                            \CFG{m} = \bot
                        }
                        {
                            ( \MD, m, \PBegin, \bot, k, \Sigma )
                            \overset{}{\longrightarrow}
                            ( \MD, m', \PBegin, pc', k, \Sigma )
                        }\\
                        & \hspace*{-8mm}\text{where }  m' = \submode{\MD}{m} \text{ and }
                                         pc' =
                                            \begin{cases}
                                                \bot ,     & \mbox{if } \CFG{m'} = \bot \\
                                                \CFGStart, & \mbox{if } \CFG{m'} \neq \bot
                                            \end{cases}  \\
        \runa{detect} &\hspace*{-8mm} \infrulesL{120}
                      {
                        \CFG{m} \ne \bot
                      }
                      {
                        ( \MD, m, \PBegin, pc, k, \Sigma \cdot \sigma )
                        \overset{}{\longrightarrow}
                        ( \MD, m, \PExecute, pc, k, \Sigma \cdot \sampling(\sigma) )
                      }\\
        \runa{execute} & \infrulesL{120}
                      {
                        \execute (\CFG{m}, pc, \sigma, \period{m}) = ( pc', \sigma' )
                      }
                      {
                        ( \MD, m, \PExecute, pc, k, \Sigma {\cdot}\sigma)
                        \overset{}{\longrightarrow}
                        ( \MD, m, \PEnd, pc', k, \Sigma' )
                      }\\
                      &
                        \begin{array}{cl}
                          \text{where } & \Sigma' = \Sigma \cdot \sigma'[ts \mapsto \sigma(ts) + \period{m}] \\
                        \end{array}   \\
        \runa{continue} & \infrulesL{120}
                      {
                        pc \neq \CFGExit
                      }
                      {
                        ( \MD, m, \PEnd, pc, k, \Sigma )
                        \overset{}{\longrightarrow}
                        ( \MD, m, \PExecute, pc, k, \Sigma )
                      }\\
        \runa{repeat} & \infrulesL{120}
                      {
                        \forall t \in \outs{\upmodes{\MD}{m}{k}} \cdot \Sigma \not\models \guard{t}
                      }
                      {
                        ( \MD, m, \PEnd, \CFGExit, k, \Sigma )
                        \overset{}{\longrightarrow}
                        ( \MD, m, \PBegin, \CFGStart, k{+}1, \Sigma )
                      }\\
        \runa{switch} &\hspace*{-8mm} \infrulesL{120}
                      {
                        \begin{array}{l}
                            \exists t \in \outs{\MD,\upmodes{\MD}{m}{k}} \cdot \Sigma \models \guard{t} \wedge \\
                          {\forall} t' {\in} \outs{\upmodes{\MD}{m}{k}} {-} \{t\} \cdot (\Sigma\not\models\guard{t'} \vee \priority{t'} {<} \priority{t})
                        \end{array}
                      }
                      {
                        ( \MD, m, \PEnd, \CFGExit, k, \Sigma )
                        \overset{}{\longrightarrow}
                        ( \MD, m', \PBegin, pc', 1, \Sigma )
                      }\\
                      & \text{where } m' = \target{t} \text{ and } pc' =
                                            \begin{cases}
                                                \bot,      & \mbox{if } \CFG{m'} = \bot \\
                                                \CFGStart, & \mbox{if } \CFG{m'} \neq \bot
                                            \end{cases} \\
                      &
    \end{array}
    \end{small}
    \]
    \caption{Operational Semantic Rules for {\modediagram}}
    \label{tbl:inferenceRules}
\end{table}

The operational rules for {\modediagram} are given in Table~\ref{tbl:inferenceRules}. Here we adopt a big-step operational semantics for {\modediagram}, which means that we only observe the start and end points of a period in the current mode, while the state changes within a period are not recorded. This is reasonable since in practice  engineers usually monitor the states at the two ends of a period to decide if it works well. In the rules, we make use of an auxiliary function $\execute$ to represent the execution results for the mode in one period.
\[
\execute: \mathcal{CFG}(V) \times \mathcal{L} \times \State \times \mathbb{R}^+ \rightarrow \mathcal{L} \times \State
\]
It receives a flow graph, a program counter, an initial state and the time permitted to execute and returns the state and program counter after the given time is expired. Its detailed definition is left in the report \cite{modechart}. We now explain the operational rules:
\begin{itemize}
    \item[1.] \runa{enter}. When the system is at the beginning of a period, if the current mode $m$ has sub-modes, the system enters the initial sub-mode of $m$.

    \item[2.] \runa{detect}. When the system is at the beginning of a period, if the current mode $m$ is a leaf mode, the system updates its state by sampling from sensors. The function $\sampling$ represents the side-effect on variables during sensor detection. The period label $l$ is changed to be $\PExecute$, indicating that the system will then perform  computational tasks specified by the control flow graph of $m$.

    \item[3.] \runa{execute}. This rule describes the behaviors of executing CFG of the leaf mode $m$. \hide{The current state $\sigma$ is stored as the last element in the trace of states $\Sigma$. }The function $\execute$ is used to compute the new state $\sigma'$ from $\sigma$. The computation task may be finished in the current period and $pc' = \CFGExit$ holds or the task is not finished and the program counter points to some location in the control flow graph. The value of  the timestamp variable $ts$  in $\sigma'$ is equal to its value in state $\sigma$ plus the period of the mode $m$.

    \item[4.] \runa{continue}. This rule tells that when the computation task in leaf mode is not finished in a period, it will continue its task in the next period. In this case, the system is implicitly not allowed to switch to other modes from the current mode. When moving to  the next period, sensor detection is skipped.

    \item[5.] \runa{repeat}. This rule specifies the behavior of restarting the flow graph when the computation task is finished in a period. When it is at the end of a period and the system finishes executing the flow graph  ($pc = \CFGExit$), if there is no transition guard enabled, the system stays in the same mode and restarts the computation specified by the flow graph.

    \item[6.] \runa{switch}. This rule specifies the behavior of the mode transition. There exists a transition $t$, whose guard holds on the sequence of states $\Sigma$. And the priority of $t$ is higher than that of any other enabled transitions.

\end{itemize}

%% file: property.tex
\section{The Property Specification Language}\label{sec:specification}

\begin{figure}[t]
  \centering
  \includegraphics[scale=0.6]{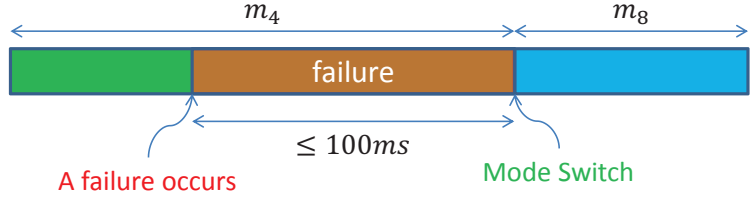}
  \caption{A Property about Failure}
  \label{fig:propertyExample}
\end{figure}

We adopt the Interval Temporal Logic (ITL)~\cite{ITLMaszkowskiM83} as the property specification language. The reason why we adopt the interval-based logic instead of state-based logics like LTL or CTL is that most of the properties the domain engineers care about are related to some duration of time. For instance, the engineers would like to check if the system specified by {\modediagram} can stay in a specific state for a continuous period of time  instead of just reaching this state. Another typical scenario illustrated in Fig.~\ref{fig:propertyExample} is that, ``\textit{when the system control is in mode $m_4$, if a failure occurs, it should switch to mode $m_8$ in 100 ms}''. The standard LTL formula $\Box(failure \wedge m_4 \Rightarrow \Diamond m_8)$ can be used to specify that ``\textit{when the system is in mode $m_4$, and a failure occurs, it should switch to mode $m_8$}''. But the real-time feature ``\textit{in 100 ms}'' is lost. Though the extensions of LTL or CTL may also describe the interval properties to some extent, it is more natural for the domain engineers to use interval-based logic since the intuitive chop operator ($^\frown$) is available in ITL.

An interval logic formula can be interpreted over a time interval \cite{IntervalCalculus} or over a ``state interval'' (a sequence of states)~\cite{ITLMaszkowskiM83} . \hide{For example,  that a formula $\phi ^\frown \psi$ holds in a time interval $[b, e]$ signifies that there exists $m$ ($b \leq m \leq e$) such that $\phi$ holds in interval $[b, m]$ and $\psi$ holds in $[m, e]$.  } As explained  later in this section, our proposed specification language will be interpreted in the latter way \cite{ITLMaszkowskiM83} except for a small modification on the interpretation of  the chop operator ($^\frown$).

\subsection{Syntax}

\begin{figure}[t]
  \centering
    $
        \begin{array}{crl}
              \textit{Terms}  &   \theta     \ \triangleq  & r \mid v \mid l \mid f(\theta_1, \ldots, \theta_n) \\
              \textit{Formulas} &   \phi, \psi \ \triangleq & \ \tvalue \mid \fvalue \mid p(\theta_1, \ldots, \theta_n)  \mid \neg \phi \mid \phi \wedge \psi \mid \phi ^\frown \psi \\
        \end{array}
    $
  \caption{The Syntax of ITL}
  \label{fig:syntaxITL}
\end{figure}

The syntax of the specification language is defined in Fig~\ref{fig:syntaxITL}, where
\begin{itemize}
    \item The set of terms $\theta$ contains real-value constants $r$, temporal variables $v$\hide{(whose values may depend on some time intervals)}, a special variable $l$, and  functions  $f(\theta_1, \ldots, \theta_n)$ (with $f$ being an $n$-arity function symbol   and $\theta_1,\ldots,\theta_n$ being terms). \hide{(1) Constant symbol $r$ denotes a real value. (2) Temporal symbol $v$ is used to represent the variable whose value depends on the given interval. (3) Function symbol $f(\theta_1, \ldots, \theta_n)$ means a n-arity real value function.}
    \item Formulae can be boolean constants  ($\tvalue$, $\fvalue$), predicates ($p(\theta_1, \ldots, \theta_n)$ with $p$, an $n$-arity predicate symbol), classical logic formulae (constructed using $\neg$, $\wedge$, etc), or interval logic formulae (constructed using $\CHOP$). \hide{(1) Constant symbols $\tvalue$, $\fvalue$ denote the boolean constants $\True$ and $\False$, respectively. (2) Predicate symbol $p(\theta_1, \ldots, \theta_n)$ means a n-arity predicate. (3) Boolean connections: negation($\neg$) and conjunction ($\wedge$) (4) Modal connection $\CHOP$ is a binary modality which \emph{chop}s an interval into two consecutive sub-intervals.} If the formula $\phi \CHOP \psi$ holds for an interval $\ell$, it means that the interval $\ell$ can be ``chopped'' into two sub-intervals, where $\phi$ holds for the first sub-interval and $\psi$  holds for  the second one.
\end{itemize}

As a kind of temporal logic, ITL also provides the $\Box$ and $\Diamond$ operators. They are defined as the abbreviations of $\CHOP$.

\[
\Diamond \phi \triangleq \tvalue \CHOP (\phi \CHOP \tvalue), \text{ for some sub-interval }, \ \
\Box \phi     \triangleq \neg \Diamond (\neg \phi), \text{ for all sub-intervals }
\]

By the specification language proposed here, we can describe the properties the domain engineers may desire. For instance, the following property describes the scenario shown in Fig.~\ref{fig:propertyExample}.

\[
    \Box(
    m_4 \wedge (\neg \failure \CHOP \failure) \CHOP \tvalue \Rightarrow
    m_4 \wedge (\neg \failure \CHOP (\failure \wedge l \le 100)) \CHOP m_8 \CHOP \tvalue
    )
\]


\begin{table}[t]
    \centering
    \begin{tabular}{l}
        $
        \begin{small}
            \begin{array}{l}
                \mathcal{I_T}(r, \Sigma) = r \\
                \mathcal{I_T}(l, \Sigma) = \begin{cases}
                                           \begin{array}{ll}
                                           \sigma_{n-1}(ts) - \sigma_{0}(ts) & \text{ if } \Sigma  = \sigma_0.\ldots.\sigma_{n{-}1} \\
                                           \infty                            & \text{ if } \mid \Sigma\mid = \infty \hide{\text{ is infinite }}
                                           \end{array}
                                           \end{cases} \\
                \mathcal{I_T}(v, \sigma_0 . \Sigma) = \sigma_0(v) \\
                \mathcal{I_T}(f(\theta_1, \ldots, \theta_n), \Sigma)  =  f\left(\mathcal{I_T}(\theta_1, \Sigma), \ldots, \mathcal{I_T}(\theta_n, \Sigma)\right) \\
            \end{array}
        \end{small}
        $ \\
        \\
        $
        \begin{small}
            \begin{array}{ccl}
                \mathcal{I_F}(p(\theta_1, \ldots, \theta_n), \Sigma) = \True & \text{ iff } &  p(\mathcal{I_T}(\theta_1, \Sigma), \ldots, \mathcal{I_T}(\theta_n, \Sigma)) \\
                 \mathcal{I_F}(\tvalue, \Sigma) = \True & \text{ iff } & \textit{always} \\
                \mathcal{I_F}(\fvalue, \Sigma) = \False & \text{ iff } & \textit{always} \\
                \mathcal{I_F}(\neg \phi, \Sigma) = \True & \text{ iff } & \mathcal{I_F}(\phi, \Sigma) = \False \\
                \mathcal{I_F}(\phi \wedge \psi, \Sigma) = \True & \text{ iff } &
                \mathcal{I_F}(\phi, \Sigma) = \True \text{ and } \mathcal{I_F}(\psi, \Sigma) = \True\\
                \mathcal{I_F}(\phi ^\frown \psi, \Sigma) = \True & \text{ iff } &
                \exists k  < \infty \cdot \ \Sigma = (\sigma_0 \ldots \sigma_k \cdot \Sigma') \wedge \\
                & & \hspace*{4mm} \mathcal{I_F}(\phi, \sigma_0\ldots\sigma_k) = \True \wedge \mathcal{I_F}(\psi, \Sigma') = \True \\
            \end{array}
        \end{small}
        $ \\
    \end{tabular}
  \caption{Interpretation of the Specification Language}\label{tbl:semanticsITL}
\end{table}

\subsection{Interpretation}

Terms/formulae in our property specification language are interpreted in the same way as in Maszkowski \cite{ITLMaszkowskiM83}, where
an interval is represented by  a finite or infinite sequence of states ($\Sigma = \sigma_0 \sigma_1 \ldots \sigma_{n-1} \ldots$), where $\sigma_i \in \State$. The interpretation is given by two functions (1) term interpretation :$\mathcal{I_T} \in  \Terms \times \Intv \mapsto \mathbb{R}$, and (2) formula interpretation function: $\mathcal{I_F} \in \Formulas \times \Intv \mapsto \{\True, \False\}$. Table~\ref{tbl:semanticsITL} defines these two functions, where $ts$  denotes the variable $\timestamp$. \hide{and a special temporal variable $l$ is introduced to denote the interval length. }The value of the variable $\timestamp$ increases with the elapse of the time. i.e.,  for any two states in the same interval $\sigma_i, \sigma_j$, if $i {<} j$, then $\sigma_i(ts) {<} \sigma_j(ts)$. Thus, we can compute the length of  time interval based on the difference of the two time stamps located in the first and last states respectively. The interpretation of a variable $v$ on $\Sigma$ is the evaluation of $v$ on the first state of $\Sigma$. Note that our chop operator requires that the first sub-interval of $\Sigma$ is restricted to be finite no matter whether the interval $\Sigma$ itself is finite or not.

%% file: verification.tex
\section{{\modediagram} Verification by Statistical Model Checking}\label{sec:statistical}

As a modelling \& verification framework for periodic control systems, {\modediagram} supports the modeling of periodic behaviors, mode transition, and complex computations involving linear or non-linear mathematical formulae. Moreover, it also provides a property specification language to help the engineers capture requirements. In this section, we will show how to verify  that an {\modediagram} model satisfies  properties formalized in the specification language. There are two main obstacles to apply classic model checking techniques on {\modediagram}: (1) {\modediagram} models involve complex computations like non-linear mathematic formulae; (2) {\modediagram} models are open systems which need intensive interactions with the outside.

Our proposed approach relies on Statistical Model Checking (SMC)~\cite{SMCSenVA04,SMCYounes05,SMCTA2011, UppaalSMC}. SMC is a simulation-based technique that runs the system to generate traces, and then uses statistical theory to analyze the traces to obtain the verification estimation of the entire system. SMC usually deals with the following quantitative aspect of the system under verification~\cite{SMCYounes05}:
\begin{itemize}
    \item [] What is the probability that a random run of a system will satisfy the given property $\phi$?
\end{itemize}

Since the SMC technique verifies the target system with the probability estimation instead of the accurate analysis, it is very effective when being applied to open and non-linear systems. Because SMC depends on the generated traces of the system under verification, we shall  briefly describe how to simulate an {\modediagram} and then present an SMC algorithm for {\modediagram}.

\subsection{{\modediagram} Simulation}

 The {\modediagram} model captures a reactive system~\cite{ReactiveHarel}. The {\modediagram} model executes and interacts with its external environment in a \emph{control loop} in one period as follows: (1) Accept inputs via sensors from the environment. (2) Perform computational tasks. (3) Generate outputs to drive other components. The {\modediagram} simulation engine simulates the process of the control loop above.

Generally speaking, the simulation is implemented according to the inference rules defined in Table~\ref{tbl:inferenceRules}. However, the behaviors of an {\modediagram} model depends not only on the {\modediagram} itself, but also on the initial state and the external environment. When we simulate the {\modediagram} model, the initial values are randomly selected from a range specified by the control engineers from CAST. As a specification language, the type of variables defined in {\modediagram} can be real number. To implement the simulation, we use float variables instead, which may introduce some problems on precision. There are lots of techniques can be adopted to check if any loss of precision may cause problems\cite{floatHigham}. Because the simulation doesn't take care of the platform to deploy the system specified by the {\modediagram}, the time during simulation is not the real time, but the logic time. For each iteration in the \emph{control loop}, the time is increased by the length of period of the current mode.

To make the simulation be executable, we have to simulate the behaviors of the environment to make the {\modediagram} model to be closed with its environment. The environment simulator involving kinematic computations designed by the control engineers is combined with the {\modediagram} to simulate the physical environment the {\modediagram} model interacts with. In the beginning of each period, the simulator checks whether there are sub-modes in the current mode. If so, the simulator takes the initial sub-mode as the new current mode. When the current mode is a leaf mode, the simulator invokes the library simulating the physical environments and updates the internal state by getting the value detected from sensors. Then the simulator executes the control flow graph in the leaf mode. We assume that there is enough time to execute the CFG. The situation that tasks are allowed not to be finished in one period is not considered during simulation. In the end of each period, the guards of transitions are checked. The satisfactions of $\duration$ and $\after$ guards do not only depend on the current state, but also the past states. The simulator sets a counter for each $\duration$/$\after$ guard instead of recording the past states. As an {\modediagram} model is usually a non-terminating periodic system, the bound of periods is set during the process of simulation.

\subsection{SMC Algorithm}

\begin{figure}[t]
    \centering
    \begin{tabular}{rl}
    \textbf{input}  & $\MD$: the {\modediagram}, $\phi$: property, $B$: bound of periods \\
                    & $\delta$: confidence, $\epsilon$: approximation \\
    \textbf{output} & $p$: the probability that $\phi$ holds on an arbitrary run of $\MD$\\
    \textbf{begin}  & \\
      10 & $N := 4 * \frac{\log{\frac{1}{\delta}}}{\epsilon^2}$, $a := 0$ \\
      20 & \textbf{for} $i := 1$ \textbf{to} $N$ \textbf{do}\\
      30 & \hspace*{4mm} generate an initial state $s_0$ randomly \\
      40 & \hspace*{4mm} simulate the {\modediagram} from $s_0$ in $B$ periods to get the state trace $\Sigma$\\
      50 & \hspace*{4mm} \textbf{if} ($\mathcal{I_F}(\phi, \Sigma) = \True$) \textbf{then} $a := a + 1$\\
      60 & \textbf{end for} \\
      70 & \textbf{return} $\frac{a}{N}$ \\
      \textbf{end} &
    \end{tabular}
    \caption{Probability Estimation for \modediagram}\label{fig:smc}
\end{figure}

We apply the methodology in ~\cite{SMCYounes05} to estimate the probability that a random run of an {\modediagram} will satisfy the given property $\phi$ with a certain precision and certain level of confidence. The statistical model checking algorithm for {\modediagram} is illustrated in Fig.~\ref{fig:smc}. Since the run of the {\modediagram} usually is infinite, the users can set the length of the sequence by the number of periods based on the concrete application. This algorithm firstly computes the number $N$ of runs  based on the formula $N := 4 * \log(1/ \delta) / \epsilon^2$ which involves the confidence interval $[p - \delta, p + \delta]$ with the confidence level $1 - \epsilon$. Then the algorithm generates the initial state (line 30) and gets a state trace $\Sigma$ by the inference rules defined in Table~\ref{tbl:inferenceRules} (line 40). The algorithm in line 50 decides whether $\phi$ holds on the constructed interval based on the interpretation for the specification language mentioned in Section~\ref{sec:specification}. If the interpretation is $\True$, the algorithm increases the number of traces on which property $\phi$ holds. Line 70 returns the probability for the satisfaction of $\phi$ on the {\modediagram}.

\subsection{Experiments}

\begin{figure}[t]
    \centering
    \includegraphics[scale=0.5]{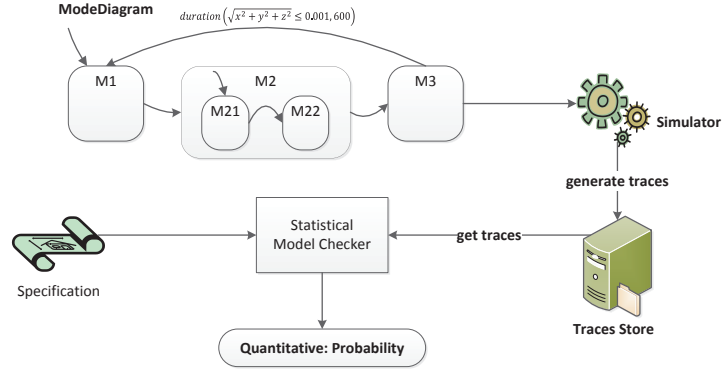}\\
    \caption{The Framework of Implementation}\label{fig:implement}
\end{figure}

We have implemented the {\modediagram} modeling and verification framework and applied it onto several real periodic control systems. The implementation framework of SMC is illustrated in Fig.~\ref{fig:implement}, where the simulator is used to simulate the {\modediagram} by the proposed operational semantics and the generated traces are for the statistical model checker. One of the real periodic control systems (termed as {\SA}) is for spacecraft control developed by Chinese Academy of Space Technology. Fig.~\ref{fig:mcoverall}(shown in Section 1) is a small portion of the {\modediagram} model for system {\SA}.


%

We communicate with the engineers in CAST, summarize several properties the two models of spacecrafts should obey, and present these properties in our specification language. A total of 12 properties are developed by the engineers and these properties are verified on the systems {\SA}. We only highlight three properties  because the verification results on these three properties reveal two defects.

\begin{itemize}
\item
\textit{After 3000 seconds, the system will eventually reach the stable state forever}
\[
\ell \ge 3000 \Rightarrow tt ^\frown \Box(\sqrt{\omega_x^2 + \omega_y^2 + \omega_z^2} \leq 0.1 \wedge \sqrt{\dot{\omega_x}^2 + \dot{\omega_y}^2 + \dot{\omega_z}^2} \leq 0.01))
\]
where $\omega_x$, $\omega_x$ and $\omega_z$ are angles. $\dot{\omega_x}$, $\dot{\omega_x}$ and $\dot{\omega_z}$ are angle rates.

\item
\textit{The system starts from mode $m0$, and then it will finally switch to mode $m5$ or $m6$ or $m8$, and stay in one of these three modes forever}
\[
(\textsf{mode} = 0) ^{\frown} tt ^{\frown} \Box(\textsf{mode} = 5 \vee \textsf{mode = 6} \vee \textsf{mode} = 8)
\]

\item
\textit{Whenever the system switches to mode $m4$ and then leaves $m4$, during its stay in $m4$, it firstly stays in
sub-mode $G0$, and then it switches to sub-mode $G1$, and then $G2$.}
\[
    \begin{array}{c}
        \Box(\textsf{mode} \neq 4 ^{\frown} \textsf{mode} = 4 ^{\frown} \textsf{mode} \neq 4 \Rightarrow \textsf{mode} \neq 4 ^{\frown}\\
        \textsf{mode} = 4 \wedge (\textsf{gm} = 0 ^{\frown} \textsf{gm} = 1 ^{\frown} \textsf{gm} = 2)^{\frown} tt)
    \end{array}
\]
\end{itemize}

For the parameters of the statistical model checking algorithm, we set the half length of confident interval to be $1\%$ ($\delta = 1\%$) and the error rate to be $5\%$ ($\epsilon = 5\%$). Based on this algorithm, the total $7369$ traces for each control system are required to be generated to compute the probabilities during the verification process.

During the verification phase by the statistical model checking on {\modediagram}, two design defects in system {\SA} are uncovered by analyzing the verification results: (1)  A variable is not initialized properly. (2) A value from sensors is detected from the wrong hardware address. In the traditional developing process in CAST, these two defects may be revealed only after a prototype of the software is developed and then tested. Our approach can find such bugs in design phase and reduce the cost to fix defects.

%% file: related.tex
\section{Related Work}\label{sec:discussion}

Our {\modediagram} can be broadly considered as a variant of Statecharts~\cite{statechart}, where a mode in  {\modediagram} is similar to a state in the Statecharts. However, we note the following distinctions: (1) In Statecharts, when a transition guard holds, the system immediately switches to the target state. But in {\modediagram}, mode switches are only allowed to be triggered at the end of a period.\hide{, since it is for the description of the periodic-driven system.} (2) In Statecharts, a transition guard is usually a boolean expression on the current(source) state; while in {\modediagram},  transition guards may involve past states via predicates like $\mathsf{during}$ and $\mathsf{after}$. (3) In Statecharts, all observations on the system are the states; while   {\modediagram} also concerns about the computation aspect of the system by means of the flow graphs provided in the leaf modes.

Timed Automata are a modeling tool for the description and  verification of real-time systems~\cite{AlurTTA,UppaalBehrmannDLPY11}. It provides the $clock$ variable to support the time explicitly. Timed Automata only focus on the linear computation for time since it has nice time zone semantics supporting the timed verification. Hybrid Automata~\cite{HybridAlurCHH92} extend the traditional automata to deal with complex computation like the difference and differentiation while it is not a systematic modeling tool which supports the rich modeling mechanisms like the hierarchy, types etc.

Giese et al.~\cite{rtsc} have proposed a semantics of real-time variant of Statecharts by introducing the Hierarchical Timed Automata. In another work~\cite{RealTimeUMLGiese} they have presented a compositional verification approach to the real time UML designs. A. K. Mok et al. have developed a kind of herarchical real-time chart named ``Modechart''~\cite{ModechartMok}. Compared with Giese et al.~\cite{rtsc}, parallel modes are supported in Modechart.

Stateflow is the Statechart-like language used in the commercial software Matlab/Simulink~\cite{stateflow}. The Stateflow language enriches  Statecharts to allow it to support  flow-based and state-based computations together for specifying  discrete event systems. Our {\modediagram}  focuses more on  periodic control systems, which can be regarded as a specific type of  discrete event systems, and it provides the first class element \textit{period} to facilitate the precise modeling of periodic-driven systems. The transitions in Stateflow can be attached with a flowchart to describe  complicated computation,  the {\modediagram} specifies the flow graph for the computation in its leaf modes. While Stateflow focuses only on the modeling aspect of the systems,  the {\modediagram} integrates modeling and reasoning by providing a property specification language with a verification algorithm.

Some researchers  introduce \textit{operational modes}~\cite{OhMultiMode,SchmitzAE05Mode} during the modeling in hardware/software co-synthesis. The operational mode is essentially a state in the automaton, but it can be attached a flowchart for the description of the computation. It does not support the nested mode and period explicitly. However, it is actually an informal modeling notation because it allows to specify the system behaviors in natural language. Our {\modediagram} is a lightweight formal notation for the modeling with its precise operational semantics.

Giotto is also a periodic-driven modeling language proposed by Henzinger et al.~\cite{giotto}. The main difference between Giotto and {\modediagram} is the computation mechanism provided. The tasks in a mode can be performed in parallel in Giotto while  the details of the tasks are omitted and are moved to the implementation stage. The {\modediagram} supports the detailed description of the computation in their leaf modes since the design of it is targeted for  control systems which may involve rich algorithms. The {\modediagram} does not support the parallel computation explicitly at present since it could bring the nondeterminism at the design level. The emphasis of the Giotto is more for the modeling and synthesis of  parallel tasks while the {\modediagram} is for the modeling and verification based on the proposed specification language.

Runtime Verification is a verification approach based on extracting information by executing the system and using the information to detect whether the observed behaviors violating the expected properties~\cite{RVHavelund,RVStolz}. The verification approach we apply in this paper is also a kind of runtime verification. But our methodology is the off-line analysis, while ~\cite{RVStolz} applies an on-line monitoring approach using Aspect-J. The reason to propose off-line analysis is that the cost to decide if an ITL formula is satisfiable on a given trace is expensive, so information extraction and analysis are separated to two phases in our approach.

\section{Conclusion}\label{sec:conclusion}

In this paper, we propose the \underline{M}ode \underline{D}iagram \underline{M}odeling framework ({\modediagram}), a domain-specific formal visual modeling language  for periodic control systems. To support formal reasoning, {\modediagram} is equipped with a property specification language based on interval temporal logic and a statistical model checking algorithm. The property specification language allows engineers to precisely capture various properties they desire, while the verification algorithm allows them to reason about {\modediagram} models with respect to those properties. The viability and effectiveness of the proposed {\modediagram} framework have been demonstrated by a number of real-life case studies, where defects of spacecraft control systems have been uncovered in the early design stage.